\documentclass[aip,pre,superscriptaddress,groupedaddress,twocolumn]{revtex4}  
\usepackage{amsfonts}
\usepackage{amsmath}
\usepackage{amssymb}
\usepackage{version}
\usepackage{color}
\usepackage{graphicx}%
\includeversion{New_connection}
\excludeversion{Old_connection}

\begin{document}
\title{Defining velocities for accurate kinetic statistics in the GJF thermostat}

\author{Niels Gr{\o}nbech-Jensen}

\affiliation{Department of Mechanical \& Aerospace Engineering,
  University of California, Davis, CA 95616, USA}

\affiliation{Department of Mathematics, University of California,
  Davis, CA 95616, USA}

\author{Oded Farago}

\affiliation{Department of Chemistry, University of Cambridge,
  Cambridge CB2 1EW, United Kingdom}

\affiliation{Department of Biomedical Engineering, Ben Gurion
  University of the Negev, Be'er Sheva, 84105 Israel}
\begin{abstract}
We expand on two previous developments in the modeling of
discrete-time Langevin systems. One is the well-documented
Gr{\o}nbech-Jensen Farago (GJF) thermostat, which has been
demonstrated to give robust and accurate configurational sampling of
the phase space. Another is the recent discovery that also kinetics
can be accurately sampled for the GJF method.  Through a
complete investigation of all possible finite difference
approximations to the velocity, we arrive at two main conclusions:~1)
It is {\it not} possible to define a so-called on-site velocity such
that kinetic temperature will be correct and independent of the time
step, and~2) there exists a set of infinitely many possibilities for
defining a two-point (leap-frog) velocity that measures kinetic
energy correctly for linear systems in addition to the correct
configurational statistics obtained from the GJF algorithm. We give
explicit expressions for the possible definitions, and we incorporate
these into convenient and practical algorithmic forms of the normal
Verlet-type algorithms along with a set of suggested criteria for
selecting a useful definition of velocity.
\end{abstract}
\maketitle
\section{Introduction}
\label{sec:intro}
Simulations of statistical properties in complex systems have been a
subject of intense interest for the past several
decades~\cite{AllenTildesley,Frenkel,Rapaport}, especially in the area
of Molecular Dynamics, where thermodynamic ensembles are sampled by
following the temporal evolution of large numbers of interacting
particles. This is done by numerically integrating Newton's equation
of motion of each particle. The most commonly used algorithm for this
purpose is the one proposed by Verlet~\cite{Verlet}, which is correct
to second order in the integration time step, $dt$, and conserves
energy in long-time integrations. The Verlet algorithm samples the
microcanonical ensemble, but the more frequently used ensemble in
statistical-mechanics is the canonical $(N,V,T)$ ensemble where the
temperature of the system, rather than its energy, is constant.  Many
methods for controlling the temperature of a simulated system
(thermostats) have been developed, and most of them fall into two
major categories: Deterministic (e.g., Nos{\'e}-Hoover
\cite{Nose,Hoover}) and stochastic (Langevin) thermostats
\cite{SS,BBK,Pastor_88,Gunsteren,Loncharich_2004,Vanden,Melchionna_2007,ML,GJF1,Paquet}.
The deterministic approach includes additional degrees of freedom,
which act as an energy reservoir and thereby mimic a thermal heat
bath. A requirement for such method is that the temperature of a
simulated system can be reliably measured in order for the system to
interact properly with the heat-bath. The stochastic approach is based
on the assumption that each particle in the system has its behavior
modeled by a Langevin equation~\cite{Langevin}
\begin{eqnarray}
m\dot{v}+\alpha\dot{r} & = & f+\beta \; , \label{eq:Langevin}
\end{eqnarray}
where $m$ is the mass of an object with spatial coordinate, $r$, and
velocity $v=\dot{r}$, and $f$ is the force acting on the
coordinate. This is Newton's second law with two additional terms
representing the interactions with a heat bath: (i) Linear (in $v$)
friction, which is represented by the friction constant $\alpha \ge
0$, and stochastic white noise, $\beta(t)$, which can be chosen to be
a Gaussian distributed variable.  These terms are thermodynamically
matched through the fluctuation-dissipation theorem by~\cite{Parisi}
\begin{eqnarray}
\langle\beta(t)\rangle & = & 0 \\
\langle\beta(t)\beta(t^\prime)\rangle & = & 2\alpha k_BT \delta(t-t^\prime) \; , 
\end{eqnarray}
where $k_B$ is Boltzmann's constant and $T$ is the thermodynamic
temperature.

Integrating numerically a Langevin equation of motion poses a
challenge since discrete time tends to distort the conjugated
relationship between the positional coordinate and its corresponding
momentum (see Appendix in Ref.~\cite{2GJ}).  A resulting problem is
that the kinetic and configurational measures of temperature disagree,
which is a concern for both the integrity of a simulation and the
extraction of self-consistent information that may depend on
configurational as well as kinetic sampling. It is therefore
imperative to understand how to properly define a kinetic measure
consistent with the statistics from the trajectory. {\color{black}We note that the
issue of defining configurational temperatures for also investigating non-equilibrium
ensembles have been extensively pursued in, e.g., Refs.~\cite{Jepps,Rickayzen,Daivis,Jackson}.
As our concern in this paper is to devise kinetically predictable velocity definitions
that accompany the GJF method {\color{black}{\it in equilibrium}}, we will not directly
be addressing the configurational temperature, since the GJF method is already
known to provide time-step independent configurational sampling in equilibrium {\color{black}\cite{GJF1,GJF2}}.}

The possibility of
creating a discrete-time simulation method that gives statistically
sound results for {\it both} configurational and kinetic measures was
first reported in Ref.~\cite{2GJ}{\color{black}, where the statistically correct 2GJ half-step
velocity was identified}, and comprehensively demonstrated to give
robust (i.e., independent of the integration time step, $dt$, for the
entire stability range of time-steps) statistics for both nonlinear
and complex molecular systems.  The algorithm, which is rooted in the
statistically sound spatial trajectory of the GJF
algorithm~\cite{GJF1,GJF2,GJF4}, is formulated in a typical {\it
  Leap-Frog} (LF) form that is easily implemented into existing
Molecular Dynamics codes. Subsequently, a related formulation of the GJF algorithm,
with similar kinetic {\color{black}averages},
has been identified~\cite{Farago}. It is the objective of this paper
to demonstrate that there exists a large set of kinetically
correct velocity definitions given by one free parameter, and that this set
includes the reported velocities \cite{2GJ,Farago}.

\section{Discrete-Time Langevin Dynamics}
\label{sec_lab_II}

Since the starting point of this work is the spatial GJF trajectory,
we give a brief review of the features of this method here. The GJF
algorithm for simulating Eq.~(\ref{eq:Langevin}) in discrete time
is~\cite{GJF1}
\begin{eqnarray}
r^{n+1} & = & r^n + b [dt\,
  v^n+\frac{dt^2}{2m}f^n+\frac{dt}{2m}\beta^{n+1}] \label{eq:gjf_r}\\
v^{n+1} & = & a\,
v^n+\frac{dt}{2m}(af^n+f^{n+1})+\frac{b}{m}\beta^{n+1} \,
, \label{eq:gjf_v}
\end{eqnarray}
where $r^n$, $v^n$, and $f^n$ are the discrete-time GJF position,
velocity, and force, respectively, at time $t_n$, and where
\begin{eqnarray}
a & = & \frac{\displaystyle{1-\frac{\alpha
      dt}{2m}}}{\displaystyle{1+\frac{\alpha dt}{2m}}} \label{eq:a}\\
b & = & \frac{\displaystyle{1}}{\displaystyle{1+\frac{\alpha
      dt}{2m}}} \,  \label{eq:b}
\end{eqnarray}
are the coefficients that define the discrete-time {\color{black}attenuation}.  The
associated discrete-time noise is
\begin{eqnarray}
\beta^{n+1} & = & \int_{t_n}^{t_{n+1}}\beta(t^\prime)\,dt^\prime \,
, \label{eq:discrete_beta}
\end{eqnarray}
which results in an uncorrelated Gaussian random variable with zero
mean and a variance given by the temperature and friction coefficient:
\begin{eqnarray}
  \langle\beta^n\rangle & = & 0 \label{eq:noise_dis_ave}\\
  \langle\beta^n\beta^l\rangle & = & 2\alpha k_BT dt \delta_{n,l} \,
, \label{eq:noise_dis_std}
\end{eqnarray}
where $\delta_{n,l}$ is the Kronecker delta function.

As was pointed out in Ref.~\cite{GJF1}, the basic thermodynamic
properties for a flat potential, $f=0$ are well behaved.
  The equipartition theorem for the kinetic energy is satisfied:
\begin{equation}
   \langle E_k\rangle  =  \frac{1}{2}m\langle(v^n)^2\rangle \; = \;
   \frac{1}{2}k_BT,
\end{equation}
and the configurational Einstein diffusion
\begin{eqnarray}
D_E & = &
\lim_{n dt\rightarrow\infty}\frac{\left\langle\left(r^{n}-r^0\right)^2
  \right\rangle}{2ndt}
\; = \; \frac{k_BT}{\alpha} \label{eq:D_E} \; ,
\end{eqnarray}
yields the correct expectation for any set of
simulation parameters, including the time step. Appendix~A
shows that also the Green-Kubo evaluation of diffusion can yield the correct
value $D_E$ if a particular Riemann approximation is applied to the Green-Kubo
integral.

Notice that the velocity attenuation factor $a$ ($|a|<1$)
  in Eq.~(\ref{eq:a}) is negative for time steps larger than
  \begin{equation}
    dt_a=2m/\alpha.
  \label{eq:dta}
\end{equation}
Choosing $dt>dt_a$ does not affect the robust configurational sampling
properties of the GJF method~\cite{GJF1}, but (as will be discussed
below) it may lead to certain nonphysical features of the discrete-time
velocity autocorrelation.

\subsection{GJF for Linear Systems, $f=-\kappa r$}

While the kinetic measure of diffusion in a flat potential can be
defined correctly for the GJF velocity variable, the 
 harmonic oscillator, given by $f^n=-\kappa r^n$ with
$\kappa>0$, shows how configurational and kinetic statistics are no
longer mutually consistent when the potential has
curvature. In~\cite{GJF1}, we showed that for
  $n\rightarrow\infty$
\begin{eqnarray}
  \langle E_p\rangle & = & \frac{1}{2}\kappa\langle(r^n)^2\rangle \; = \;
  \frac{1}{2}k_BT\label{eq:gjf_epot}\\
  \langle E_k\rangle & = & \frac{1}{2}m\langle(v^n)^2\rangle \; = \;
  \frac{1}{2}k_BT\left(1-\frac{(\Omega_0dt)^2}{4}\right) \; ,\label{eq:gjf_ekin}
\end{eqnarray}
where $\Omega_0=\sqrt{\kappa/m}$ is the natural frequency of the
oscillator. These results hold for any time step smaller than the
Verlet stability limit $\Omega_0dt\leq 2$. The appealing features of
the GJF algorithm is given by Eqs.~(\ref{eq:D_E}) and
(\ref{eq:gjf_epot}) as these indicate sound results for
configurational statistics, which is the aim of computer simulation
studies of, e.g., molecular systems at equilibrium. {\color{black}As promoted in Ref.~\cite{2GJ},
t}he discrete-time velocity
is predominantly an auxiliary variable, {\color{black} which for simulations
of equilibrium is} used primarily for
assessing the temperature of the simulated system via the mean kinetic
energy.  Since $\Omega_0$ is an expression of the curvature of the
potential, Eq.~(\ref{eq:gjf_ekin}) shows that a general system cannot
be simulated with a correct kinetic statistical measure using the
GJF velocity Eq.~(\ref{eq:gjf_v}).

With the useful GJF spatial trajectory and the accompanying depressed
on-site GJF velocity, $v^n$, which results in imperfect kinetic
statistics, we here investigate the kinetic
response of {\it all} finite difference velocities.

\subsection{A general finite difference velocity}

Since the aim of this section is to explore velocity definitions that
may accompany the GJF trajectory, it is natural to write the GJF
method in its St{\o}rmer-Verlet form~\cite{GJF2,2GJ}:
\begin{eqnarray}
  r^{n+1} & = & 2br^n-ar^{n-1}+\frac{b\,dt^2}{m}f^n+
  \frac{b\,dt}{2m}(\beta^n+\beta^{n+1}) \, , \nonumber \\ \label{eq:gjf_sv}
\end{eqnarray}
with the GJF velocity Eq.~(\ref{eq:gjf_v}) expressed as~\cite{2GJ}
\begin{eqnarray}
  v^n & = & \frac{r^{n+1}-(1-a)r^n-ar^{n-1}}{2\, b \, dt}
  +\frac{\beta^n-\beta^{n+1}}{4m} \, . \label{eq:gjf_v2}
\end{eqnarray}
Inspired by Eq.~(\ref{eq:gjf_v2}), we define a velocity
$w$ in the general finite-difference form
\begin{eqnarray}
  w & = & \frac{\gamma_1r^{n+1}+\gamma_2r^n+\gamma_3r^{n-1}}{dt}
  +\frac{\gamma_4\beta^n+\gamma_5\beta^{n+1}}{m} \; , \nonumber
  \label{eq:GJF-2GJF}\\
\end{eqnarray}
where $\gamma_i$ are unit-less constants that are to be determined,
and where the two noise terms, $\beta^n$ and $\beta^{n+1}$, span the
time interval of the finite difference, $t_{n-1}<t<t_{n+1}$ [see
  Eq.~(\ref{eq:discrete_beta})]. Notice that we have not attached a
superscript on $w$ that indicates at which time this velocity is
represented, since this general expression is representing any
velocity approximation in the interval spanned by the finite difference.
Specifically, we recognize the usual three-point on-site and two-point half-step
velocities in the frictionless ($\alpha=0$) Verlet algorithm
\begin{eqnarray}
v^n & = & \frac{r^{n+1}-r^{n-1}}{2\,dt} \label{eq:on-site}\\
v^{n+\frac{1}{2}} & = & \frac{r^{n+1}-r^{n}}{dt} \, ,\label{eq:half-step}
\end{eqnarray}
for $\gamma_1=-\gamma_3=\frac{1}{2}$, $\gamma_2=\gamma_4=\gamma_5=0$
and $\gamma_1=-\gamma_2=1$, $\gamma_3=\gamma_4=\gamma_5=0$,
respectively. We also recognize the GJF velocity,
  $v^n$, in Eq.~(\ref{eq:gjf_v2}), as represented by $\gamma_1=1/2b$,
$\gamma_2=-(1-a)/2b$, $\gamma_3=-a/2b$, $\gamma_4=-\gamma_5=1/4$.

We start by writing the most basic statistical requirement to a
velocity variable, namely
\begin{eqnarray}
\langle w w\rangle & = & \frac{k_BT}{m}\, . \label{eq:Kin_Stat}
\end{eqnarray}
Using Eqs.~(\ref{eq:gjf_sv}) and (\ref{eq:GJF-2GJF}), $\langle
w w\rangle$ can be rewritten
\begin{eqnarray}
\langle w w\rangle & = &
\frac{\gamma_1^2+\gamma_2^2+\gamma_3^2}{dt^2}\langle
r^nr^n\rangle+2\frac{\gamma_1\gamma_2+\gamma_2\gamma_3}{dt^2}
\langle r^nr^{n+1}\rangle \nonumber \\
& + & 2\frac{\gamma_1\gamma_3}{dt^2}\langle r^{n-1}r^{n+1}\rangle +
\frac{\gamma_4^2+\gamma_5^2}{m^2}\langle\beta^n\beta^n\rangle\nonumber\\
& + & 2\frac{\gamma_1\gamma_4}{m\,dt}\langle
r^{n+1}\beta^n\rangle+2\frac{\gamma_1\gamma_5+\gamma_2\gamma_4}{m\,dt}\langle
r^n\beta^n\rangle\; . \label{eq:w2_1}
\end{eqnarray}
From Eqs.~(\ref{eq:noise_dis_std}), (\ref{eq:gjf_epot}), and
(\ref{eq:gjf_sv}) we obtain the relevant correlations:
\begin{eqnarray} 
  \langle\beta^nr^n\rangle & = &
  (b-a)k_BT\,dt\label{eq:correlation1}\\
  \langle\beta^nr^{n+1}\rangle & = &
  (2+a-b\Omega_0^2dt^2)(b-a)k_BT\,dt \\
  \langle r^nr^{n+1}\rangle & = &
  \frac{k_BT}{\kappa}\left(1-b\frac{\Omega_0^2dt^2}{2}\right) \\
  \langle r^{n-1}r^{n+1}\rangle & = &
  \frac{k_BT}{\kappa}\left(1-b(b+1)\Omega_0^2dt^2+b^2\frac{\Omega_0^4dt^4}{2}
  \right)\, ,\nonumber\\
  \ \label{eq:correlation4}
\end{eqnarray}
which, when inserted into Eq.~(\ref{eq:w2_1}), yield
\begin{eqnarray}
&& \frac{m\langle w w\rangle}{k_BT} \; = \;
  \frac{\gamma_1^2+\gamma_2^2+\gamma_3^2+2(\gamma_1\gamma_2+\gamma_2\gamma_3
    +\gamma_3\gamma_1)}{\Omega_0^2dt^2}\nonumber\\
  &&-b[\gamma_1\gamma_2+\gamma_2\gamma_3+2(1+b)\gamma_3\gamma_1]
  +4(\gamma_4^2+\gamma_5^2)\frac{1-b}{b}\nonumber \\
  &&+4(1-b)[\gamma_1\gamma_5+\gamma_2\gamma_4+(1+2b)\gamma_1\gamma_4]\nonumber\\
  && + \Omega_0^2dt^2b\gamma_1[b\gamma_3-4(1-b)\gamma_4]\,
  . \label{eq:w2_2}
\end{eqnarray}
While this expression is somewhat cumbersome, it immediately reveals
key information about possible definitions of kinetically robust
velocities to accompany the GJF trajectory.  First, from the
requirement that Eq.~(\ref{eq:Kin_Stat}) is satisfied for any (stable)
$dt$, it follows that the terms in Eq.~(\ref{eq:w2_2}) proportional to
both $(\Omega_0dt)^{-2}$ and $(\Omega_0dt)^2$ must be zero. Thus, we
must require that
\begin{eqnarray}
  0 & = & \gamma_1^2+\gamma_2^2+\gamma_3^2+2(\gamma_1\gamma_2
  +\gamma_2\gamma_3+\gamma_3\gamma_1), \label{eq:cond_Wdtinv}\\
0 & = & b\gamma_1[b\gamma_3-4(1-b)\gamma_4].  \label{eq:gam_34}
\end{eqnarray}
Second, since $w$ represents the velocity during the time interval
$t_n\le t<t_{n+1}$, we will further require that
$\gamma_1\neq0$. Moreover, in the limit $\alpha dt\rightarrow0$
($a,b\rightarrow1$, $\beta^n=\beta^{n+1}=0$) the coefficient
$\gamma_1$ should become either $\gamma_1\rightarrow\frac{1}{2}$ or
$\gamma_1\rightarrow1$, such that $w$ becomes one of the two known
velocities given in Eqs.~(\ref{eq:on-site}) and (\ref{eq:half-step}) in
that limit. Under these conditions, Eq.~(\ref{eq:gam_34}) yields the
following noise term associated with $\beta^n$:
\begin{eqnarray}
  \gamma_4\beta^n = \frac{b}{1-b}\gamma_3\sqrt{2\alpha k_BT\,dt}\,\sigma^n
  =2m\gamma_3 \sqrt{2\frac{k_BT}{\alpha dt}} \, \sigma^n, \label{eq:cond_34_2}
\end{eqnarray}
where $\sigma^n\in N(0,1)$ is a random number with a
  standard normal distribution. The requirement for
  $\gamma_4$ to be confined is that $\gamma_3\rightarrow0$ faster
than $(\alpha dt)^\frac{1}{2}$ for $\alpha dt\rightarrow0$. This
condition, however, cannot be met by on-site velocity variables $v^n$, which
in the limit $\alpha dt\rightarrow0$ must coincide with
Eq.~(\ref{eq:on-site}), where $\gamma_1=-\gamma_3=\frac{1}{2}\neq0$,
and the limit $\gamma_3\rightarrow-\frac{1}{2}$ would create a
diverging noise term in any on-site velocity definition as $\alpha
dt\rightarrow0$. We therefore conclude that {\it no reasonable on-site
  finite-difference velocity that has correct and time step
  independent kinetic statistics can be defined} such that it will
approach the expected central difference approximation in the limit
$\alpha dt\rightarrow0$.

\subsection{{\color{black}Two-point} velocity, $\gamma_3=\gamma_4=0$}

{\color{black}In light of Eq.~(\ref{eq:half-step}), we will throughout this paper denote a two-point
velocity $w=u^{n+\frac{1}{2}}$ as one given by Eq.~(\ref{eq:GJF-2GJF})
with $\gamma_3=\gamma_4=0$, such that the value of the denoted two-point
velocity pertains to the time interval $t_n<t<t_{n+1}$.}

Setting $\gamma_3=\gamma_4=0$, considerately simplifies Eq. (\ref{eq:w2_2}) to
\begin{eqnarray}
  && \frac{m\langle (u^{n+\frac{1}{2}})^2\rangle}{k_BT} \; =
  \; \frac{(\gamma_1+\gamma_2)^2}{\Omega_0^2dt^2}\nonumber \\
&&-b\gamma_1\gamma_2 +4\gamma_5^2\frac{1-b}{b}
 +4(1-b)\gamma_1\gamma_5 \, . \label{eq:w2_3}
\end{eqnarray}
With the requirement that Eq.~(\ref{eq:Kin_Stat}) holds
  for any $dt$, we have $\gamma_1=-\gamma_2$, which then yields the
condition:
\begin{eqnarray}
b\gamma_1^2+4\frac{1-b}{b}\gamma_5^2+4(1-b)\gamma_1\gamma_5 & = & 1\,
. \label{eq:cond_15}
\end{eqnarray}
From this expression, we can determine $\gamma_5$ as a function of a
given $\gamma_1$:
\begin{eqnarray}
  \gamma_5 & = &-\frac{1}{2}b\gamma_1\pm\frac{1}{2}
  \sqrt{b\frac{1-b^2\gamma_1^2}{1-b}}, \label{eq:gam5_gam1}
\end{eqnarray}
which implies that $(b\gamma_1)^2\le 1$.  We will also
require that $\gamma_1\rightarrow1$ for $b\rightarrow1$ in order to
recover the corresponding velocity in the limit
of either continuous time or no friction, $\alpha
  dt\rightarrow 0$ [see discussion above, after
    Eq.~(\ref{eq:gam_34})]. Thus, we have identified a
family of velocities that yield the correct average
kinetic energy in discrete-time:
\begin{eqnarray}
u^{n+\frac{1}{2}} & = &
\gamma_1\frac{r^{n+1}-r^n}{dt}+\frac{\gamma_5}{m}\beta^{n+1}\, ,
\label{eq:2GJF}
\end{eqnarray}
where $\gamma_5$ is determined by the parameter $\gamma_1$, which is
limited in magnitude by $|b\gamma_1|\le1$. 

Using Eq.~(\ref{eq:2GJF}) for the velocity,
together with Eq.~(\ref{eq:gjf_sv}) for the GJF trajectory, we arrive
for $\gamma_1\neq0$ at the following general LF GJF algorithm:
\begin{eqnarray}
u^{n+\frac{1}{2}} & = & au^{n-\frac{1}{2}} +\gamma_1\frac{bdt}{m}f^n
+\frac{\Gamma_4}{2m}\beta^n+\frac{\Gamma_5}{2m}\beta^{n+1}\label{eq:GJF-LF-u}\\
r^{n+1} & = & r^n+\frac{dt}{\gamma_1}\,u^{n+\frac{1}{2}}
-\frac{\gamma_5}{\gamma_1}\frac{dt}{m}\beta^{n+1},\label{eq:GJF-LF-r}
\end{eqnarray}
where
\begin{eqnarray}
\Gamma_4 & = & b\gamma_1-2a\gamma_5\label{eq:Gamma_1}\\
\Gamma_5 & = & b\gamma_1+2\gamma_5\label{eq:Gamma_2}\, .
\end{eqnarray}
The general scheme Eqs.~(\ref{eq:GJF-LF-u})-(\ref{eq:Gamma_2}) can be
also written in the following form, involving both the denoted {\color{black}two-point}
velocity, $u^{n+1/2}$, and the on-site velocity, $v^n$, expressed in
Eqs.~(\ref{eq:gjf_v}) and (\ref{eq:gjf_v2}):
\begin{eqnarray}
u^{n+\frac{1}{2}} & = &
b\gamma_1\left[v^n+\frac{1}{2m}\frac{\Gamma_5}{b\gamma_1}\beta^{n+1}
  +\frac{dt}{2m}f^n
  \right] \label{eq:GJF2_u} \\
r^{n+1} & = & r^n
+\frac{dt}{\gamma_1}\,u^{n+\frac{1}{2}}
-\frac{\gamma_5}{\gamma_1}\frac{dt}{m}\beta^{n+1}\label{eq:GJF2_r}\\
v^{n+1} & = &\frac{a}{b\gamma_1}u^{n+\frac{1}{2}}
+\frac{1}{2m}\frac{\Gamma_4}{b\gamma_1}\beta^{n+1}
+\frac{dt}{2m}f^{n+1} \; . \label{eq:GJF2_v}
\end{eqnarray}
This compact form of the method further illuminates the meaning of the
parameter $\gamma_1$ beyond the direct scaling of the finite-difference 
{\color{black}two-point} velocity, observed in
Eq.~(\ref{eq:2GJF}). As mentioned
at the beginning of section~\ref{sec_lab_II}, the total attenuation
factor of the velocity over one time step is $a$, and this factor is
shown in Eqs.~(\ref{eq:GJF2_u})-(\ref{eq:GJF2_v}) to be partitioned
into two parts: The first is the attenuation $b\gamma_1$ of the
velocity $v^n$ into the velocity $u^{n+\frac{1}{2}}$; the
second is the attenuation $a/b\gamma_1$ of the
velocity $u^{n+\frac{1}{2}}$ into $v^{n+1}$. The product of the two
attenuation factors is obviously $a$. It is physically reasonable to expect that
the attenuation factor is positive and not larger than unity in either
of the two parts of the time step. A negative attenuation factor
$b\gamma_1$ implies a peculiar velocity that is in
directional opposition to the surrounding trajectory, whereas a factor
which is greater than unity implies velocity amplification. Thus, in
order to have a physically meaningful description of the velocity
attenuation, we must choose (i) $a\geq 0$ [$dt\leq dt_a$ - see
  discussion around Eq.~(\ref{eq:dta})], and (ii) $a\leq b\gamma_1
\leq 1$.  With that said, we reemphasize that any velocity
defined by Eq.~(\ref{eq:2GJF}) will always yield the correct average of the
kinetic energy, and that this form only requires that $|b\gamma_1|\leq
1$.

\subsection{Special cases}

We now highlight
the following three choices of $\gamma_1$ as {\it examples} of
velocity definitions:\\

\noindent
{\bf Case A: $\sqrt{b}\gamma_1=1$, $\gamma_5=0$}. This is the {\color{black} 2GJ half-step}
 velocity given in Ref.~\cite{2GJ},
\begin{eqnarray}
u^{n+\frac{1}{2}}_{A} & = & \frac{r^{n+1}-r^{n}}{\sqrt{b}\,  dt}\, ,
\label{eq:u_A}
\end{eqnarray}
and it is the {\it optimal} amplitude rescaling $\gamma_1$ of the
standard definition Eq.~(\ref{eq:half-step}), since it is the only form where $\gamma_5=0$; i.e.,
the only form where the {\color{black}symmetry of the central difference is not broken by an additional noise
contribution in order to yield the correct kinetic energy}. The coefficients to
the noise terms in Eq.~(\ref{eq:GJF-LF-u}) are given by
$\Gamma_4=\Gamma_5=\sqrt{b}$.  \\

\noindent
{\bf Case B: $b\gamma_1=1$, $\gamma_5=-\frac{1}{2}$}. This
velocity is given in Ref.~\cite{Farago}:
\begin{eqnarray}
u^{n+\frac{1}{2}}_{B} & = & \frac{r^{n+1}-r^{n}}{b\, dt}
-\frac{1}{2m}\beta^{n+1}\, , \label{eq:u_B}
\end{eqnarray}
and is the {\it maximum} amplitude rescaling $\gamma_1$ of the standard
definition Eq.~(\ref{eq:half-step}). It
includes an explicit noise contribution in order to achieve the
correct kinetic energy statistics.
The velocity attenuation is here assigned
exclusively to the second half of the time step as seen from Eqs.~(\ref{eq:GJF2_u})-(\ref{eq:GJF2_v}).
The coefficients to the noise terms in Eq.~(\ref{eq:GJF-LF-u}) are
given by $\Gamma_4=2b$ and $\Gamma_5=0$.\\

\noindent
{\bf Case C: $\gamma_1=1$,
  $\gamma_5=-\frac{1}{2}(b-\sqrt{b(1+b)})$}. This case defines the
velocity from the {\it neutral} amplitude rescaling $\gamma_1=1$ of the
standard definition Eq.~(\ref{eq:half-step}), which is
\begin{eqnarray}
u^{n+\frac{1}{2}}_{C} & = & \frac{r^{n+1}-r^{n}}{dt}
-\frac{1}{2m}(b-\sqrt{b(1+b)})\,\beta^{n+1} \, . \label{eq:u_C}
\end{eqnarray}
While this expression has a non-trivial pre-factor
  $\gamma_5$ to the compensating noise term, {\color{black}it} may be attractive by the absence of amplitude scaling of the velocity
  Eq.~(\ref{eq:half-step}). {\color{black}Thus, t}he average velocity is {\color{black}here} correctly
representing a ballistic (constant velocity) trajectory.  The
coefficients to the noise terms in Eq.~(\ref{eq:GJF-LF-u}) are given
by $\Gamma_4=2b^2-a\sqrt{b(1+b)}$ and $\Gamma_5=\sqrt{b(1+b)}$.\\

We re-emphasize that simply obtaining the correct kinetic energy is
{\it not} a sufficient criterion for a physically reasonable definition of a
kinetically sound velocity.  As an extreme limiting example, we
highlight\\

\noindent
{\bf Case D: $\gamma_1=0$,
  $\gamma_5=\pm\frac{1}{2}\sqrt{\frac{b}{1-b}}$}. In this case
\begin{eqnarray}
  u^{n+\frac{1}{2}}_D & = & \pm\frac{1}{\sqrt{2\alpha \,m\,dt}}\,\beta^{n+1}.
  \label{eq:extreme_vel}
\end{eqnarray}
This definition produces the correct kinetic energy, but is clearly
not an appropriate definition of a meaningful velocity since it is
void of any information about the associated trajectory
$r^n$. Instead, at each time step, a random value is chosen from the
Maxwell-Boltzmann Gaussian distribution, and simply assigned to the
velocity variable.

Given that the velocity definitions of this paper are all built on the
GJF trajectory, we retain the configurational Einstein diffusion
result of Eq.~(\ref{eq:D_E}) for any of the above choices of
velocities. The corresponding Green-Kubo calculations for $f=0$
using the derived {\color{black}two-point} velocities can be found in Appendix~B.
The results show that the discrete-time Riemann sums allow for
correct, and time step independent diffusion results if the right-Riemann
sum is chosen for Case A (as also derived in Ref.~\cite{2GJ}), and if
the trapezoidal sum is chosen for Case B. A Green-Kubo expression
for Case C  also exists, but it is not given
 by one of the three traditional discrete-time Riemann sums.

We also note that the evaluation of a correct Green-Kubo
value for diffusion in a flat potential $f\equiv0$ is neither a
guarantee for correct Green-Kubo results in systems where $f\neq0$,
nor is it necessarily a good indicator of the quality of kinetic
measures for curved potentials.
For example, the GJF on-site velocity, $v^n$ (\ref{eq:gjf_v}),
produces the correct Green-Kubo result [see
    Eq.~(\ref{eq:GK_GJF})]; yet, it also produces a depressed kinetic
energy (\ref{eq:gjf_ekin}). In general, it is the Einstein definition
Eq.~(\ref{eq:D_E}) that determines the actual diffusion, since this
expression is a configurational measure for the actual square distance
an object has been displaced over a given time. The simple Green-Kubo
results shown in the Appendices are merely indicators of consistency
between kinetic and configurational properties of a
  freely diffusing particle.\\
  
\section{Numerical Simulations}
In order to validate the kinetic features obtained for the
velocities presented in this paper, we conduct
the same kind of molecular simulations that was used to validate Case A {\color{black}along with showing detailed comparisons with the well-known BBK \cite{BBK} and PBS \cite{Pastor_88} methods} in
Ref.~\cite{2GJ}. The system
consists of $N=864$ atoms, each with mass $m$, in a cubic simulation
cell with volume $V$ and periodic boundary conditions at a normalized
pressure of approximately 0.1. The interaction potential $E_{p}(r)$ is
a {\color{black}splined, short range Lennard-Jones potential given by
\begin{widetext}
\begin{eqnarray}
\frac{E_p(|r|)}{E_0} & = &
\left\{\begin{array}{lcl} \displaystyle{\left(\frac{|r|}{r_0}\right)^{-12}-2\left(\frac{|r|}{r_0}\right)^{-6}} & , & 0<|r|\le r_s
\\ \displaystyle{\frac{a_4}{E_0}\left(|r|-r_c\right)^4+\frac{a_8}{E_0}\left(|r|-r_c\right)^8} & , & r_s<|r|<r_c \\ \displaystyle{0} & , & r_c \le |r|
\end{array}\right.
\label{eq:Eq_LJ_spline}
\end{eqnarray}
\end{widetext}
$r$ being a coordinate between a pair of particles. The spline parameters are given by
\begin{eqnarray}
\frac{r_s}{r_0} & = & \left(\frac{13}{7}\right)^{1/6} \; \approx \; 1.108683 \\
\frac{r_c}{r_0} & = & \frac{r_s}{r_0}-\frac{32E_p(r_s)}{11E_p^\prime(r_s)r_0} \; \approx \; 1.959794
\end{eqnarray}
\begin{eqnarray}
a_4 & = & \frac{8E_p(r_s)+(r_c-r_s)E_p^\prime(r_s)}{4(r_c-r_s)^4} \\
a_8 & = & -\frac{4E_p(r_s)+(r_c-r_s)E_p^\prime(r_s)}{4(r_c-r_s)^8} .
\end{eqnarray}
The potential $E_p(r)$ has minimum $-E_0$ at $|r|=r_0$, and is smoothly splined
between the inflection point of the Lennard-Jones potential and zero with continuity through the second derivative
at $|r|=r_s$, and continuity through the third derivative at $|r|=r_c$.}

\begin{figure}[t]
\centering
\scalebox{0.45}{\centering \includegraphics[trim={2cm 2.5cm 1.5cm 7.5cm},clip]{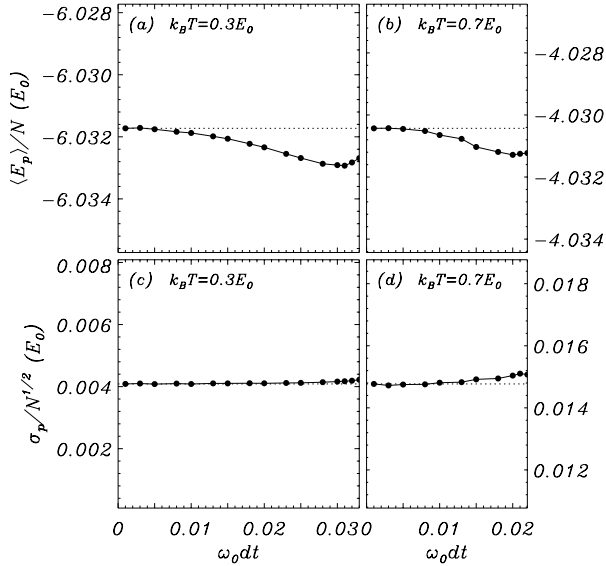}}
\caption{Statistical averages of potential energy $\langle E_p\rangle$ {\color{black}(from Eq.~(\ref{eq:Ep_ave}))}
  (a) and (b), and its standard deviation $\sigma_p$ {\color{black}(from Eq.~(\ref{eq:sig_p}))} (c) and (d) as a
  function of reduced time step $\omega_0dt$ for
  $\alpha=1\,m\omega_0$.
  (a) and (c) show results for a
  crystalline fcc state at $k_BT=0.3E_0$;
  (b) and (d) show results for a liquid state at $k_BT=0.7E_0$.
  Horizontal dotted lines indicate the
  results for small $\omega_0dt=0.001$.  }
\label{fig_1}
\end{figure}
\begin{figure}[t]
\centering
\scalebox{0.45}{\centering \includegraphics[trim={2.0cm 2.5cm 1.5cm 7.5cm},clip]{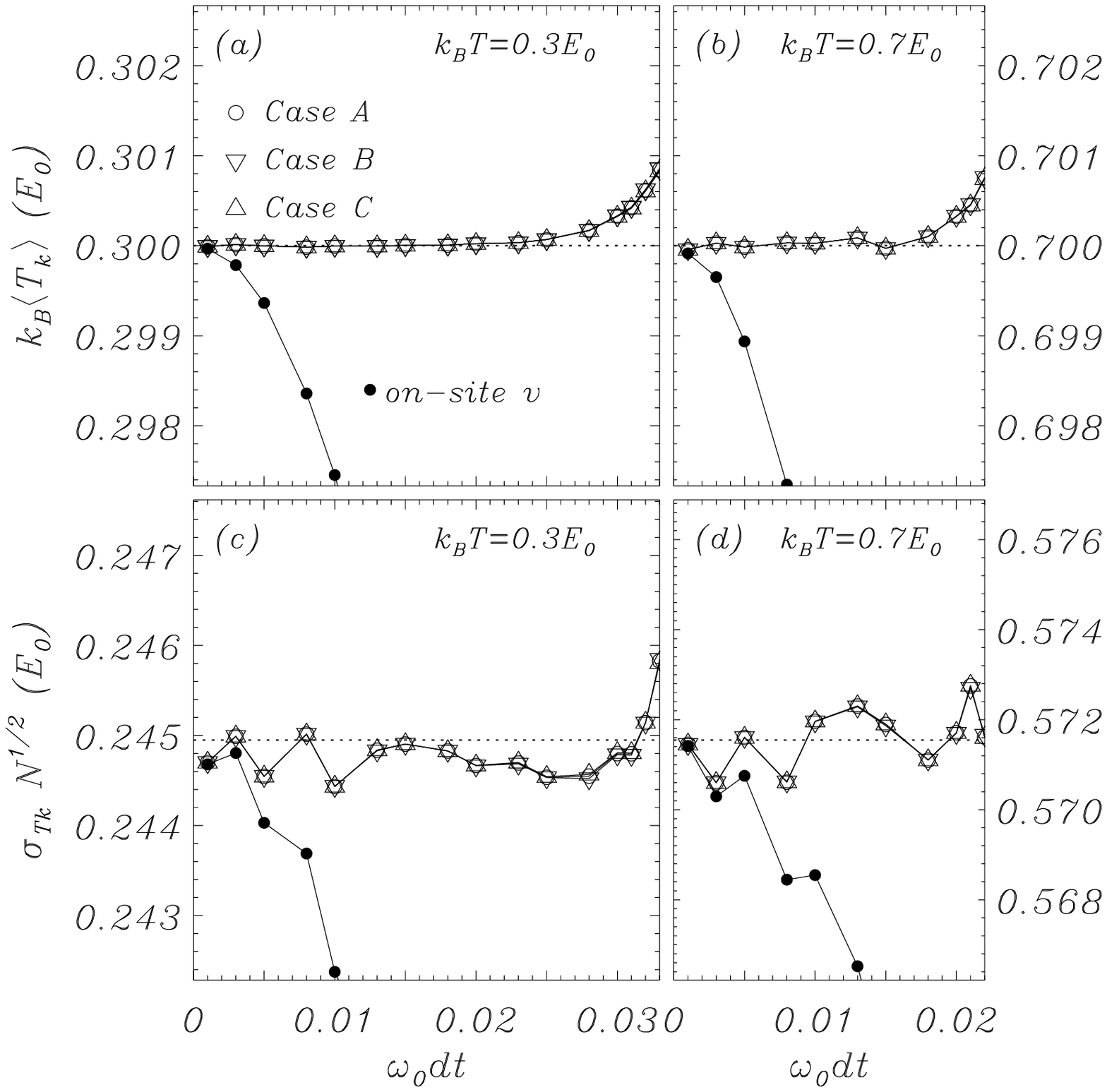}}
\caption{{\color{black}Statistical averages of the kinetic temperature (from Eq.~(\ref{eq:Ek_ave}))}
  (a) and (b), and {\color{black}their standard deviations (from Eq.~(\ref{eq:sig_k}))} (c) and (d) as a
  function of reduced time step $\omega_0dt$ for
  $\alpha=1\,m\omega_0$.
  (a) and (c) show results for a crystalline fcc state at $k_BT=0.3E_0$;
  (b) and (d) show results for a liquid state at $k_BT=0.7E_0$.
  Horizontal dotted lines indicate the
  exact, continuous time results {\color{black}, $\langle T_k\rangle^*$ and $\sigma_{T_k}^*$, given by Eqs.~(\ref{eq:Ek_exp}) and (\ref{eq:sig_Ek})}. Results shown for four velocity
  definitions: GJF on-site $v$, and velocities of Cases A, B, C.}
\label{fig_2}
\end{figure}
\begin{figure}[t]
\centering
\scalebox{0.45}{\centering \includegraphics[trim={2cm 2.5cm 1.5cm 7.5cm},clip]{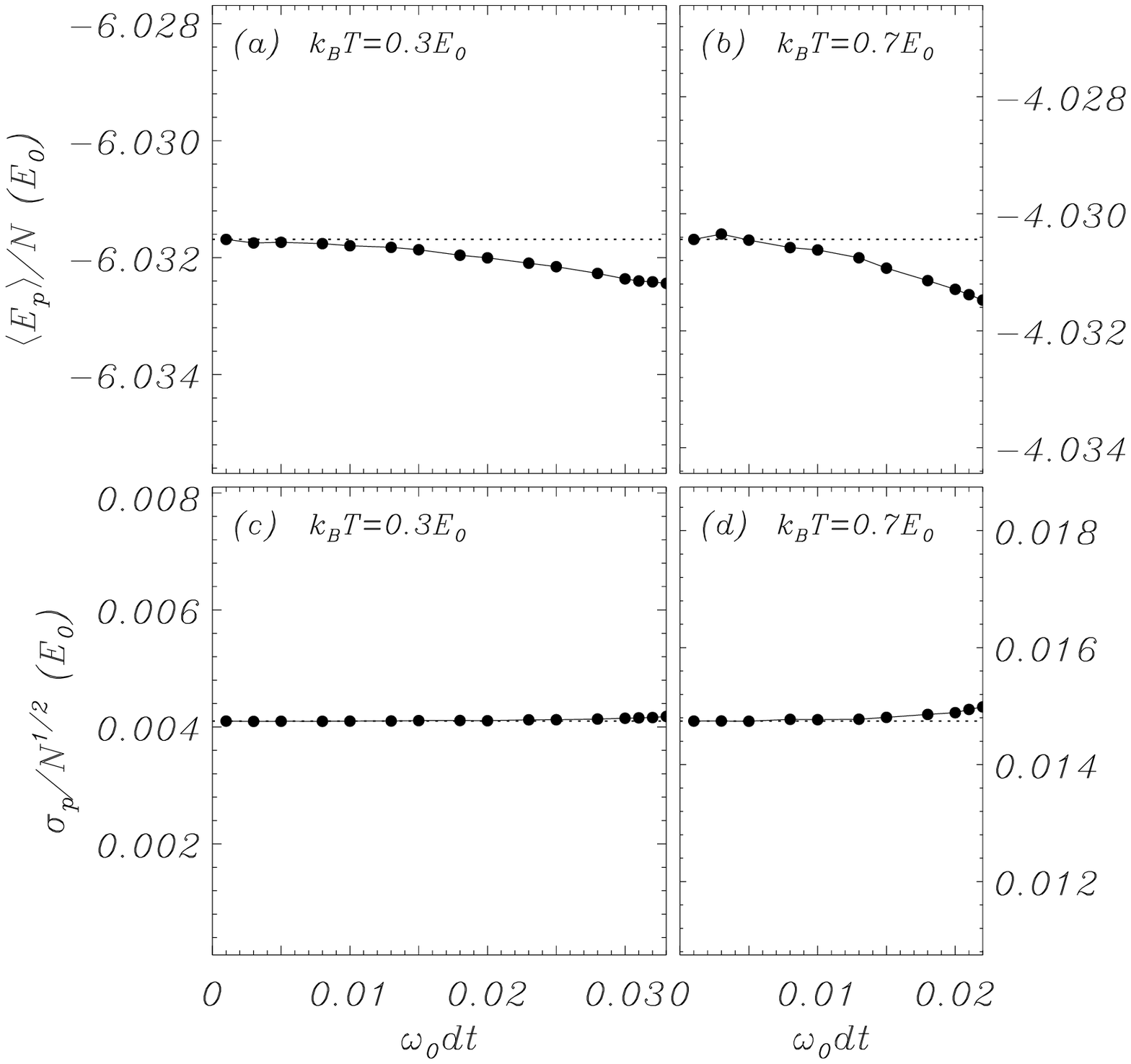}}
\caption{Statistical averages of potential energy $\langle E_p\rangle$ {\color{black}(from Eq.~(\ref{eq:Ep_ave}))}
  (a) and (b), and its standard deviation $\sigma_p$ {\color{black}(from Eq.~(\ref{eq:sig_p}))} (c) and (d) as a
  function of reduced time step $\omega_0dt$ for
  $\alpha=10\,m\omega_0$.
  (a) and (c) show results for a crystalline fcc state at $k_BT=0.3E_0$;
  (b) and (d) show results for a liquid state at $k_BT=0.7E_0$.
  Horizontal dotted lines indicate the
  results for small $\omega_0dt=0.001$.}
\label{fig_3}
\end{figure}
\begin{figure}[t]
\centering
\scalebox{0.45}{\centering \includegraphics[trim={2.0cm 2.5cm 1.5cm 7.5cm},clip]{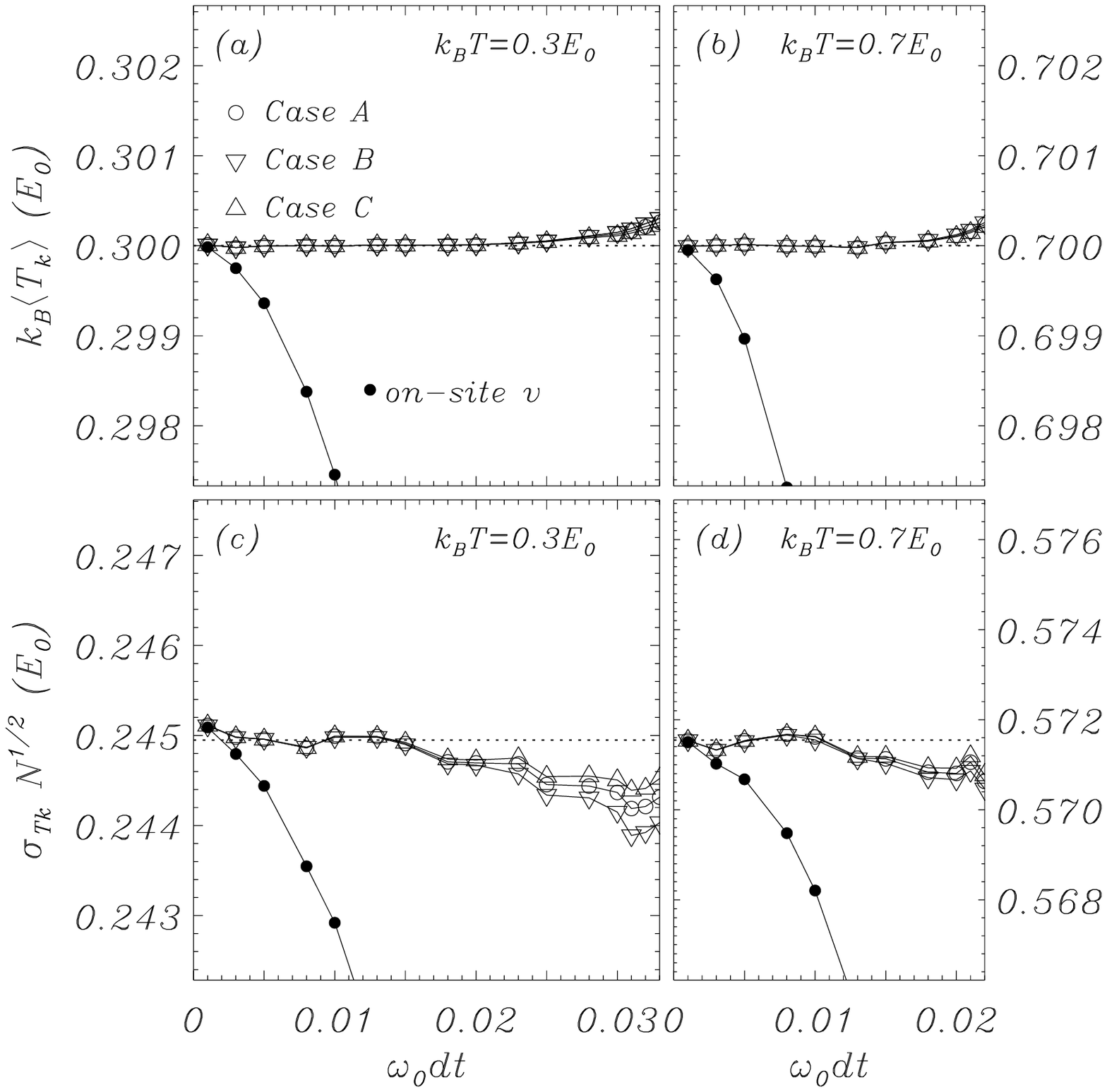}}
\caption{{\color{black}Statistical averages of the kinetic temperature (from Eq.~(\ref{eq:Ek_ave}))}
  (a) and (b), and {\color{black}their standard deviations (from Eq.~(\ref{eq:sig_k}))} (c) and (d) as a
  function of reduced time step $\omega_0dt$ for
  $\alpha=10\,m\omega_0$.
  (a) and (c) show results for a crystalline fcc state at $k_BT=0.3E_0$;
  (b) and (d) show results for a liquid state at $k_BT=0.7E_0$.
  Horizontal dotted lines indicate the
  exact, continuous time results {\color{black}, $\langle T_k\rangle^*$ and $\sigma_{T_k}^*$, given by Eqs.~(\ref{eq:Ek_exp}) and (\ref{eq:sig_Ek})}. Results shown for four velocity
  definitions: GJF on-site $v$, and velocities of Cases A, B, C.}
\label{fig_4}
\end{figure}
\begin{figure}[t]
\centering
\scalebox{0.45}{\centering \includegraphics[trim={2cm 2.5cm 1.5cm 7.5cm},clip]{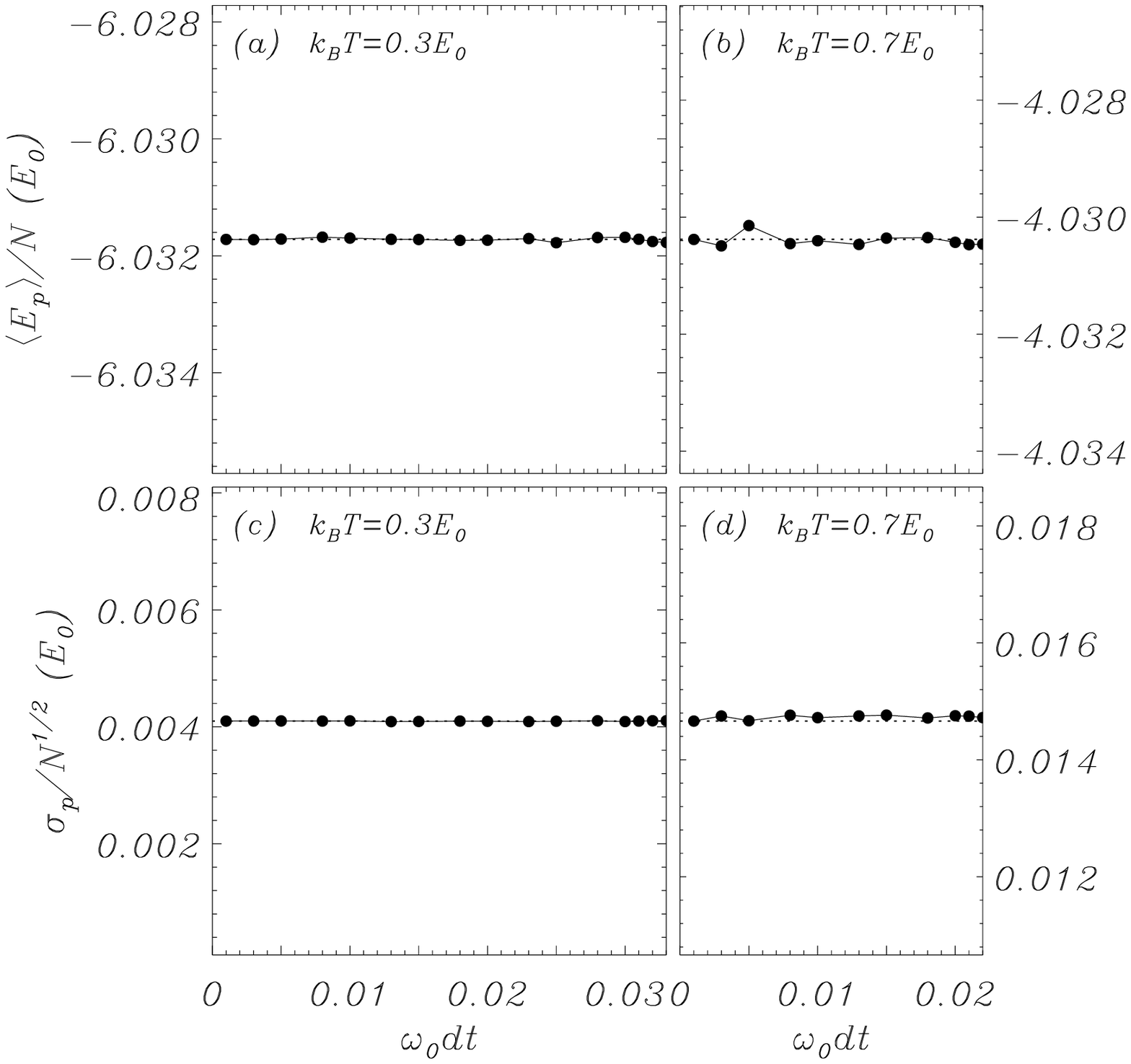}}
\caption{Statistical averages of potential energy $\langle E_p\rangle$ {\color{black}(from Eq.~(\ref{eq:Ep_ave}))}
  (a) and (b), and its standard deviation $\sigma_p$ {\color{black}(from Eq.~(\ref{eq:sig_p}))} (c) and (d) as a
  function of reduced time step $\omega_0dt$ for
  $\alpha=100\,m\omega_0$.
  (a) and (c) show results for a crystalline fcc state at $k_BT=0.3E_0$;
  (b) and (d) show results for a liquid state at $k_BT=0.7E_0$.
  Horizontal dotted lines indicate the
  results for small $\omega_0dt=0.001$.  }
\label{fig_5}
\end{figure}
\begin{figure}[t]
\centering
\scalebox{0.45}{\centering \includegraphics[trim={2.0cm 2.5cm 1.5cm 7.5cm},clip]{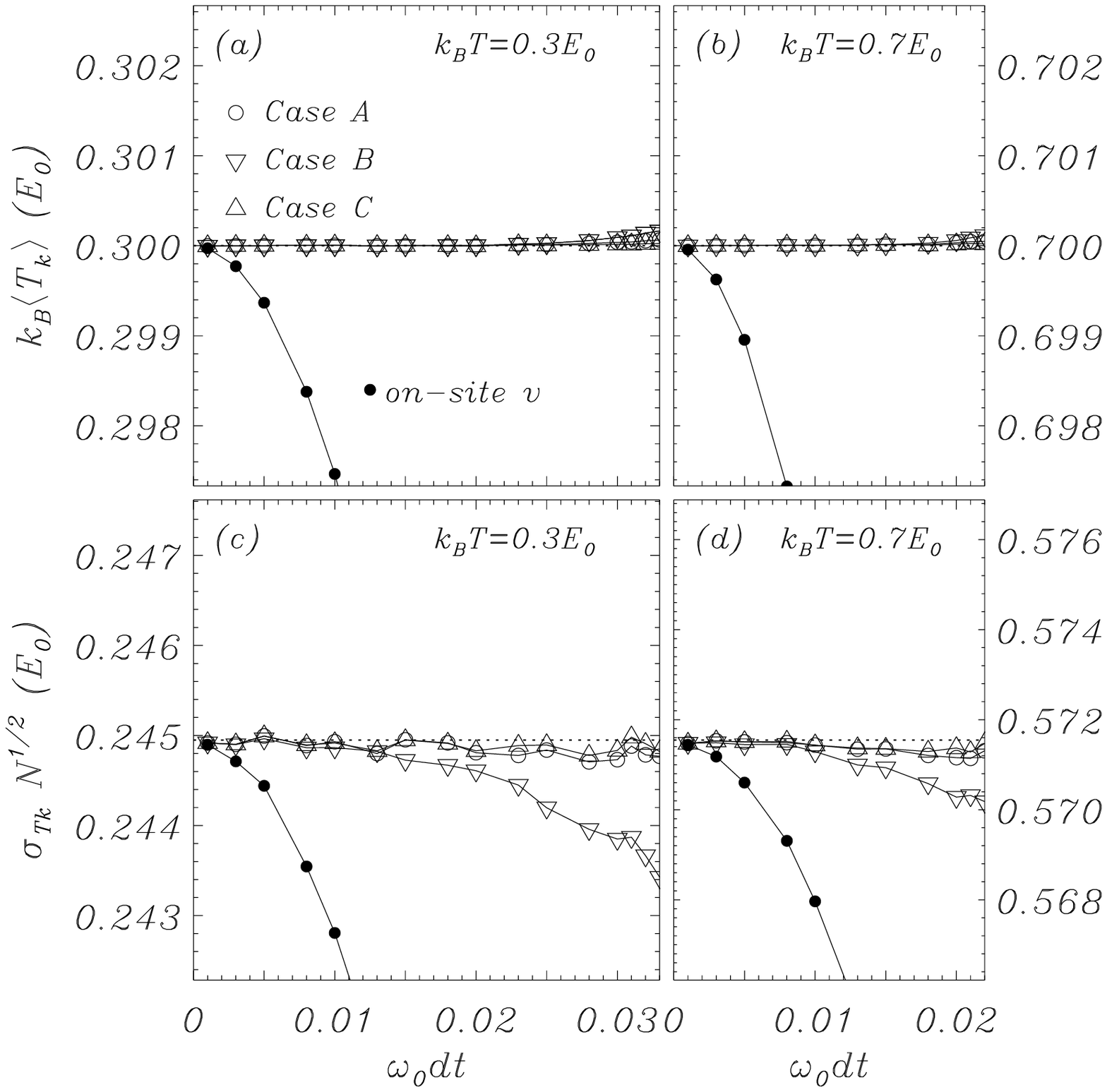}}
\caption{{\color{black}Statistical averages of the kinetic temperature (from Eq.~(\ref{eq:Ek_ave}))}
  (a) and (b), and {\color{black}their standard deviations (from Eq.~(\ref{eq:sig_k}))} (c) and (d) as a
  function of reduced time step $\omega_0dt$ for
  $\alpha=100\,m\omega_0$.
  (a) and (c) show results for a crystalline fcc state at $k_BT=0.3E_0$;
  (b) and (d) show results for a liquid state at $k_BT=0.7E_0$.
  Horizontal dotted lines indicate the
  exact, continuous time results {\color{black}, $\langle T_k\rangle^*$ and $\sigma_{T_k}^*$, given by Eqs.~(\ref{eq:Ek_exp}) and (\ref{eq:sig_Ek})}. Results shown for four velocity
  definitions: GJF- on-site $v$, and velocities of Cases A, B, C.}
\label{fig_6}
\end{figure}

With these
characteristic parameters, time is normalized to the inverse of the
characteristic frequency $\omega_0=\sqrt{E_0/mr_0^2}$.  Two
characteristically different temperatures are tested; $k_BT=0.3E_0$,
which results in a stable fcc (face centered cubic) crystal at a
volume of $V=617.2558r_0^3$, and $k_BT=0.7E_0$, which results in a
liquid state at volume $V=824.9801r_0^3$. We show results for three
different friction coefficients: $\alpha=1\, m\omega_0$, $\alpha=10\,
m\omega_0$, and $\alpha=100\, m\omega_0$. Notice that while these
values seem high at first glance, they should be compared to the
friction
$\alpha_0=m\sqrt{E_{p}^{\prime\prime}(r_0)/m}\approx\sqrt{60}m\omega_0$
relevant to the simulated density and characteristic collision
distances. From this comparison we conclude that the simulated
friction coefficients represent damping values ranging from
underdamped oscillatory to slightly overdamped behavior.

For each simulated temperature, friction value, and time step, we
calculate statistical averages over one trajectory of reduced time of
$\omega_0\Delta t=2\times10^{5}$, after the system has equilibrated
for at least the same time before statistics is acquired.
Data for all four velocities, the on-site GJF velocity
(\ref{eq:gjf_v2}), and Cases~A (\ref{eq:u_A}), B~(\ref{eq:u_B}), and
C~(\ref{eq:u_C}), are accumulated such that the displayed kinetic
results for the different velocities on Figures~\ref{fig_2},
\ref{fig_4}, and \ref{fig_6} can be directly compared. The
corresponding acquisition of the configurational statistics is shown
on Figures~\ref{fig_1}, \ref{fig_3}, and \ref{fig_5}.

As expected from previous investigations {\color{black}\cite{GJF1,2GJ,GJF2,GJF4,Finkelstein,GJ}}
of the GJF method, the
configurational statistics is excellent {\color{black}(see Figures \ref{fig_1}, \ref{fig_3}, and \ref{fig_5})}. {\color{black} The displayed quantities are derived from the potential energy $E_p^n$ of the $N$-particle system
\begin{eqnarray}
E_p^{n} & = & \sum_{i=1}^N\sum_{j>i}^N E_p(|r_i^n-r_j^n|)\, , \label{eq:Ep}
\end{eqnarray}
from which we calculate the average
\begin{eqnarray}
\left\langle E_p\right\rangle & = & \left\langle E_p^{n}\right\rangle\,  \label{eq:Ep_ave}
\end{eqnarray}
and the temporal fluctuations $\sigma_p$
\begin{eqnarray}
\sigma_p & = & \sqrt{\left\langle\left(E_p^{n}\right)^2\right\rangle-\left\langle E_p^{n}\right\rangle^2} \, . \label{eq:sig_p}
\end{eqnarray}
It is apparent that the calculated quantities are, with very good approximation, time step independent in the
entire range of stability, }with a slight decreasing
trend for increasing time steps in the average of the potential
energy, and a slight increase in its fluctuations. These deviations
are most prominent for low friction values, but are minor on a
relative scale. {\color{black}Since linear analysis of the method shows that no configurational sampling
should deviate from expectations at any stable time step (see Eq.~(\ref{eq:gjf_epot})), it follows that the observed deviations are a result of
the nonlinearities in the interaction potential, and not a result of inherent systematic algorithmic inaccuracies \cite{Ghost}.}
{\color{black}Notice that the
configurational behavior is independent of the specific velocity one
may associate with the GJF method. Thus, the data shown in Figs.~\ref{fig_1}, \ref{fig_3}, and \ref{fig_5}
represent the use of any velocity, including Cases A, B, and C.}

The
kinetic measures{\color{black}, which are the point} of interest to this presentation {\color{black}(see Figures \ref{fig_2}, \ref{fig_4}, and \ref{fig_6}),} display excellent
time step independence, as one would expect from the analysis above since these investigated velocities are engineered to
produce time step independence in their calculated kinetic
energy. {\color{black} Specifically, we investigate the kinetic energy $E_k^{n+\frac{1}{2}}$ and temperature $T_k^{n+\frac{1}{2}}$ of the $N$-particle system at time $t_{n+\frac{1}{2}}$:
\begin{eqnarray}
E_k^{n+\frac{1}{2}} & = & \frac{1}{2}\sum_{i=1}^Nm(u^{n+\frac{1}{2}})^2 \; = \; \frac{3N}{2}k_BT_k^{n+\frac{1}{2}}\, , \label{eq:Ek}
\end{eqnarray}
for which the temporal averages are expected to have the relationship
\begin{eqnarray}
\left\langle E_k\right\rangle & = & \left\langle E_k^{n+\frac{1}{2}}\right\rangle \; = \; \frac{3N}{2}k_B\langle T_k^{n+\frac{1}{2}}\rangle\; = \; \frac{3N}{2}k_B\langle T_k\rangle \nonumber \\ \label{eq:Ek_ave}
\end{eqnarray}
are expected to have the continuous-time
\begin{eqnarray}
\left\langle E_k\right\rangle^* & = & \frac{3N}{2}k_B\langle T_k\rangle^* \; = \; \frac{3N}{2}k_BT\, . \label{eq:Ek_exp}
\end{eqnarray}
We also calculate the temporal fluctuations $\sigma_{E_k}$ iand $\sigma_{T_k}$ in $E_k^{n+\frac{1}{2}}$ and $T_k^{n+\frac{1}{2}}$, respectively:
\begin{eqnarray}
\sigma_{E_k} & = & \sqrt{\left\langle\left(E_k^{n+\frac{1}{2}}\right)^2\right\rangle-\left\langle E_k^{n+\frac{1}{2}}\right\rangle^2} \; = \; \frac{3N}{2}k_B\,\sigma_{T_k} \, , \nonumber \\ \label{eq:sig_k}
\end{eqnarray}
where, in continous time, these fluctuations are expected to yield
\begin{eqnarray}
\sigma_{E_k}^* & = & \frac{3N}{2}k_B\,\sigma_{T_k}^* \; = \; \sqrt{\frac{3}{2}}\,k_BT\,\sqrt{N}\,. \label{eq:sig_Ek}
\end{eqnarray}
The simulated quantities for the three highlighted cases of velocities, Cases A, B, and C, are
shown in Figs.~\ref{fig_2}, \ref{fig_4}, and \ref{fig_6} as markers (Eqs.~(\ref{eq:Ek}) and (\ref{eq:sig_k})) and dotted horizontal lines (Eqs.~(\ref{eq:Ek_ave}) and (\ref{eq:sig_Ek})) for reference. We also show the comparable results when using the on-site GJF velocity $v^n$.
The overall impression of the simulation results for the kinetic measures is that the two-point velocities behave largely in agreement with the analytical expectations in the entire stability range of the simulations.
}
This is true for both simulated temperatures and states of
matter. It is noticeable that Cases A, B, and C behave nearly
identically, except for the
high friction value, where Case B (the maximally scaled velocity) deviates
from the two other definitions in the fluctuation measure.
However, we notice that this deviation is minor. {\color{black}It is here important to emphasize that the
kinetic results for the different velocity definitions shown in Figs.~\ref{fig_2}, \ref{fig_4}, and \ref{fig_6} are
derived from the same simulation. This explains why the fluctuations in, e.g., Fig.~\ref{fig_2}d, are identical
for Cases A, B, and C, since for low friction the three definitions are nearly identical. This also means that
the deviations between the behavior of the three two-point velocities seen in, e.g., Fig.~\ref{fig_6}c are
a true reflection of differences between the velocities.}
 We have further validated that reasonable choices of
  $\gamma_1<1$ also produce reliable results. Specifically, the cases
  $b\gamma_1=\sqrt{|a|}$ (the case for which velocity attenuation is equally
  partitioned over the two half time steps -- see Eq.~(\ref{eq:GJF2_u})-(\ref{eq:GJF2_v})),
  and $b\gamma_1=|a|$ (the case
  for which velocity attenuation is exclusively assigned to the first half of the time step)
  both yield results nearly indistinguishable from Cases A and C. We have omitted the display
  of these results in the figures for visual simplicity.
The on-site GJF
velocity is shown for comparison, and it is clear that in
  contrast to the 
velocities highlighted in this paper, the deviation for
the on-site
velocity is much more pronounced and, moreover, the error increases
  with the integration time-step.

\section{Discussion}
\label{sec_lab_IV}

Inspired by the discovery of a velocity definition that
can produce accurate kinetic statistics in conjunction with the GJF
thermostat, we have here analyzed all possible finite difference velocity
definitions that may accompany the GJF trajectory.
We draw two important conclusions: First, that it is
{\it not} possible to identify a meaningful on-site velocity such that
the kinetic measures of thermodynamics can be time step independent.
Second, that there exist an infinite number of {\color{black}two-point}
velocities such that the kinetic energy is correctly evaluated.
Having identified this family of  velocities, we
  have included them in the GJF formalism and introduced
  the general LF GJF algorithm, which is the leap-frog form
  Eqs.~(\ref{eq:GJF-LF-u})-(\ref{eq:Gamma_2}).
We have additionally written the set of methods in a convenient and
compact form of
  Eqs.~(\ref{eq:GJF2_u})-(\ref{eq:GJF2_v}), that includes any of the
defined velocities together with the native GJF on-site
velocity such that the method is entirely contained with operations
pertaining only to a single time step. 
The set of kinetically sound velocities is parameterized by a
  single parameter ($\gamma_1$), and we have highlighted three
  choices that seem either mathematically or physically attractive
  within the physical limitations to values of $\gamma_1$.

Molecular simulations of
Lennard-Jones test cases have confirmed the predicted
features of the new set of velocities, which seem to display very good
time step independent behavior throughout the stability ranges for the
time step. All three highlighted velocity definitions (which are for $\gamma_1\ge1$)
show near identical statistical behavior, except for the most
amplitude-distorted velocity, which exhibits some minor deviations in
its fluctuations for large time steps. Additionally, we have verified that also
two other reasonable choices for $\gamma_1<1$ exhibit 
results similar to the cases highlighted in the figures.

It is our hope that the complete set of defined {\color{black}GJF} velocities
will be further explored such that a more complete understanding of
the different definitions can be developed and refined for a variety
of applications. We have specifically validated
a select few of the possible velocity definitions,
  but there may very well be other choices that are more appealing in
some instances. {\color{black} The cases of immediate interest are Cases  A and C.
Case A
is the true half-step velocity that is correctly
cross correlated with the spatial degree of freedom \cite{GJ},
but does not give correct drift velocity. Case C
gives correct drift velocity, but is not a half-step velocity, since it is not
symmetrically evaluated relative to the spatial coordinate $r^n$. {\color{black}These features suggest} that
Case A should be preferred for equilibrium simulations with diffusive transport, while Case C
could be preferred for non-equilibrium simulations involving drift or ballistic transport.}

{\color{black}Our} simulations demonstrated robustness of all
  highlighted definitions in standard Molecular Dynamics systems. The many
  possibilities must be explored by the community that conduct
Langevin dynamics simulations
so that the breadth of applications, limitations, and conditions can be adequately
tested.

\appendix

\section{Green-Kubo diffusion using the GJF on-site velocity $v^n$ for $f=0$}
\label{appndxA}

The Green-Kubo equivalent of the Einstein expression for diffusion
Eq.~(\ref{eq:D_E}) is in discrete-time calculated from the
autocorrelation $\langle v^{q+n}v^q\rangle_q$ of the GJF velocity
Eq.~(\ref{eq:gjf_v}), where
\begin{eqnarray}
v^q & = & a^q v^0+\frac{b}{m}\sum_{k=1}^{q}a^{q-k}\beta^k \; .\label{eq:vq}
\end{eqnarray}
Assuming that $v^q$ is a well-equilibrated velocity
  ($|a|^q\rightarrow0$), the velocity autocorrelation in
  discrete-time reads
\begin{eqnarray}
\left\langle v^{q+n}v^q\right\rangle & = &
\frac{b^2}{m^2}\left\langle\sum_{k=1}^{q+n}a^{q+n-k}\beta^k\sum_{l=1}^qa^{q-l}
\beta^l\right\rangle  \nonumber \\
& &\xrightarrow{q\rightarrow\infty}\ 
a^n\frac{k_BT}{m} \, .\label{eq:gjf_v_cor}
\end{eqnarray}
The Green-Kubo diffusion
coefficient, $D_{GK}$, is evaluated by the continuous-time
expression~\cite{GreenKubo}
\begin{eqnarray}
D_{GK} & = & \int_{0}^{\infty}\langle
v(t+s)v(t)\rangle_{t}\,ds \label{eq:GK_def}\, .
\end{eqnarray}
However, as pointed out in Ref.~\cite{2GJ}, this definition is
somewhat ambiguous in discrete time since
different Riemann sums can be legitimately considered as
approximations to the integral in (\ref{eq:GK_def}) for $dt>0$. The
result for the trapezoidal sum is:
\begin{eqnarray}
  D_{GK} & = & \sum_{k=0}^\infty\left\langle v^{q+k}v^q\right\rangle\,dt=
  \frac{k_BT}{m}\,dt\,\left(\frac{1}{2}+\sum_{k=1}^\infty a^k\right)
  \nonumber \\
  &=&\,\frac{k_BT}{\alpha} \, .
 \label{eq:GK_GJF}
\end{eqnarray}
Thus, if the Green-Kubo integral is approximated by the trapezoidal
Riemann sum, then we obtain the correct diffusion coefficient $D_{GK}$
consistent with the correct configurational value, $D_E$, given by
Eq.~(\ref{eq:D_E}). However, we could have used other Riemann
approximations that converge to the correct results $D_E$ for
$dt\rightarrow0$, but do not exactly match it for $dt>0$. For example,
the left-Riemann sum gives $D_{GK}=D_E/b>D_E$, while the right-Riemann
sum gives $D_{GK}=D_E(a/b)<D_E$. Notice how sensitive the choice of
the discrete approximation to the integral is by comparing the
right-Riemann sum, which can yield even a nonsensical negative
diffusion constant for $dt>dt_a$ (i.e., $a<0$), to the
trapezoidal approximation, which in this case is both time step
independent and correct.

\section{Green-Kubo diffusion using a {\color{black}two-point} velocity $u^{n+\frac{1}{2}}$ for $f=0$}
\label{appndxB}

For evaluating the Green-Kubo expression (\ref{eq:GK_def}) using the
derived {\color{black}two-point} velocity, we will need the discrete-time velocity
autocorrelation function. For that purpose, we define an initial
condition at $t_{\frac{1}{2}}$, which we insert into
Eq.~(\ref{eq:GJF-LF-u}) for $f=0$. Iterating the equation $q$ times
yields
\begin{eqnarray}
  u^{q+\frac{1}{2}} & = & a^qu^{\frac{1}{2}}+a^{q-1}
  \frac{\Gamma_4}{2m}\beta^1+\frac{\Gamma_5}{2m}\beta^{q+1}\nonumber \\
  &&+\sum_{k=0}^{q-2}(\frac{\Gamma_4}{2m}
  +a\frac{\Gamma_5}{2m})a^k\beta^{q-k} \, .\label{eq:GJF-u_stoc}
\end{eqnarray}
After lengthy, but trivial, calculations, we arrive at the compact
form for the velocity autocorrelation in the limit
$q\rightarrow\infty$ (i.e., after the memory of the initial velocity
is lost)
\begin{eqnarray}
\langle u^{q+\frac{1}{2}+n}u^{q+\frac{1}{2}}\rangle & = &
\frac{k_BT}{m}\left[a^n+a^{n-1}\Gamma_4\Gamma_5\frac{\alpha
    dt}{2m}(1-\delta_{n,0})\right]\, .\nonumber \\ \label{eq:uu_auto}
\end{eqnarray}
This expression can be used to approximate the Green-Kubo expression
for diffusion by a Riemann sum. As was discussed in
  Ref.~\cite{2GJ} and Appendix~\ref{appndxA}, we can justify any Riemann sum of the
Green-Kubo integral in Eq.~(\ref{eq:GK_def}) with a discretization of
$dt$. The three obvious choices, left-Riemann $D_{lR}$, trapezoidal
$D_{tr}$, and right-Riemann $D_{rR}$ sums yield
\begin{eqnarray}
D_{lR} & = &
\frac{k_BT}{m}dt\left[1+\frac{1}{1-a}\left(a+\Gamma_4\Gamma_5\frac{\alpha
  dt}{2m}\right)\right],\label{eq:D_rR}\\
D_{tr} & = &
\frac{k_BT}{m}dt\left[\frac{1}{2}+
  \frac{1}{1-a}\left(a+\Gamma_4\Gamma_5\frac{\alpha
  dt}{2m}\right)\right]\label{eq:D_tr},
\\ D_{rR} & = &
\frac{k_BT}{m}dt\frac{1}{1-a}\left(a+\Gamma_4\Gamma_5\frac{\alpha
  dt}{2m}\right),\label{eq:D_lR}
\end{eqnarray}
where $D_{rR}\le D_{tr}\le D_{lR}$.  When selecting $\gamma_1$ and
$\gamma_5$, an additional consideration in determining if a
velocity is meaningful may be that the corresponding Green-Kubo
diffusion expressions can yield a result such that
$D_E\in[D_{rR},D_{lR}]$.  Not all choices of $\gamma_1$ yield a velocity variable satisfying
this criterion.
Therefore, we define a parameter $0\le{\cal C}\le1$, which can
determine if a given velocity (as defined by $\gamma_1$ and
$\gamma_5$) can produce $D_E\in[D_{rR},D_{lR}]$, i.e., if there exists
a ${\cal C}\in[0,1]$ for the given method such that
\begin{eqnarray}
D_E & = & \frac{k_BT}{\alpha} \; = \; D_{rR}+{\cal
  C}\frac{k_BT}{m}\,dt \, . \label{eq:cond_diff}
\end{eqnarray}
From Eq.~(\ref{eq:D_rR}) the value of ${\cal C}$ for which
Eq.~(\ref{eq:cond_diff}) is true is given by
\begin{eqnarray}
{\cal C} & = & \frac{1}{2}-\frac{\Gamma_4\Gamma_5}{2b}\,
, \label{eq:value_C}
\end{eqnarray}
and the condition for a given velocity to have a Green-Kubo diffusion
value within the acceptable range is then
\begin{eqnarray}
-b & \le & \Gamma_4\Gamma_5 \; \le \; b \, . \label{eq:cond_Gamma}
\end{eqnarray}
It is straightforward to verify that 
the above highlighted cases A, B, and C satisfy this
criterion for any value $0\le b\le1$. 
  Specifically, the optimally scaled velocity,
$u^{n+1/2}_{A}$~(\ref{eq:u_A}), gives the correct Einstein diffusion
for $D_{rR}$ (as described in Ref.~\cite{2GJ}), and the maximally
scaled velocity, $u^{n+1/2}_{B}$~(\ref{eq:u_B}), gives the correct
diffusion for $D_{tr}$.

\begin{thebibliography}{99} 

\bibitem{AllenTildesley} M.~P~. Allen, D.~J.~Tildesley, {\it Computer
  Simulation of Liquids}, Oxford University Press, Inc., New York,
  1989.

\bibitem{Frenkel} D.~Frenkel and B.~Smit, {\it Understanding Molecular
  Simulations: From Algorithms to Applications}, (Academic Press, San
  Diego, 2002).

\bibitem{Rapaport} D.~C.~Rapaport, {\it The Art of Molecular Dynamics
  Simulations}, (Cambridge University Press, Cambridge, 2004).

\bibitem{Verlet} L.~Verlet, Phys. Rev. {\bf 159}, 98 (1967).

\bibitem{Nose} S.~Nos\'{e},
J. Chem. Phys. {\bf 81}, 511 (1984).
  
\bibitem{Hoover} W. ~G.~Hoover,
Phys. Rev. A {\bf 31}, 1695 (1985).
  
\bibitem{SS} T.~Schneider and E.~Stoll,
Phys. Rev. B {\bf 17}, 1302 (1978).
  
\bibitem{BBK} A.~Br\"{u}nger, C.~L.~Brooks, and M.~Karplus,
Chem. Phys. Lett. {\bf 105}, 495 (1984).
  
\bibitem{Pastor_88} R.~W.~Pastor, B.~R.~Brooks, and A.~Szabo,
  Mol.~Phys.~{\bf 65}, 1409 (1988).
  
\bibitem{Gunsteren} W.~F.~van Gunsteren, H.~J.~C. Berendsen,
  Mol. Simul. {\bf 1}, 173, (1988).
  
\bibitem{Loncharich_2004} R.~J.~Loncharich, B.~R.~Brooks, and
  R.~W.~Pastor,
  Biopolymers {\bf 32}, 523 (1992).
  
\bibitem{Vanden}E.~Vanden-Eijnden, G.~Ciccotti,
Chem. Phys. Lett. {\bf   429}, 310 (2006).
  
\bibitem{Melchionna_2007} S.~Melchionna,
J.~Chem.~Phys.\ {\bf 127},  044108 (2007).
  
\bibitem{ML} B.~Leimkuhler, C.~Matthews,
Appl. Math. Res. Express {\bf
  2013}, 34 (2012).
  
\bibitem{GJF1} N.~Gr{\o}nbech-Jensen and O.~Farago,
Mol. Phys. {\bf  111}, 983 (2013).
  
\bibitem{Paquet} E.~Paquet, H.~L.~Viktor,
BioMed Res. Int., 183918 (2015).

\bibitem{Langevin} P.~Langevin, 
C.~R.~Acad. Sci. Paris {\bf 146}, 530 (1908).

\bibitem{Parisi} G.~Parisi, {\it Statistical Field Theory},
  (Addison-Wesley, Menlo Park, 1988).

\bibitem{2GJ} L.~F.~G.~Jensen and N. Gr{\o}nbech-Jensen,
  Mol.~Phys.~{\bf 117}, 2511 (2019).
  {\color{black}
  \bibitem{Jepps} O. G. Jepps, G. Ayton, and D.
J. Evans, Phys. Rev. E {\bf 62}, 4757 (2000).

\bibitem{Rickayzen}G. Rickayzena and J. G. Powles, J. Chem. Phys. {\bf 114}, 4333 (2001).

\bibitem{Daivis}P. J. Davis, B.A. Dalton, and T. Morishita, Phys. Rev. E {\bf 86}, 056707 (2012).

\bibitem{Jackson}N. Jackson, J. Miguel Rubi, and Fernando Bresme, Mol.~Sim.~{\bf 42}, 1214 (2016).
}
\bibitem{GJF2} N.~Gr{\o}nbech-Jensen, N.~R.~Hayre, and O.~Farago,
  Comput. Phys. Commun. {\bf 185}, 524 (2014).

\bibitem{GJF4} E.~Arad, O.~Farago, and N.~Gr{\o}nbech-Jensen,
  Isr. J. Chem. {\bf 56}, 629 (2016).

 \bibitem{Farago} O.~Farago, 
 Physica A {\bf 534}, 122210 (2019).
 
 {\color{black}
\bibitem{Ghost} L.~F.~G.~Jensen and N. Gr{\o}nbech-Jensen,
  Comput.\ Phys.\ Commun.\ (2019), DOI: 10.1016/j.cpc.2019.107011. arXiv:1902.02338.
\bibitem{Finkelstein}J.~Finkelstein, G.~Fiorin, and B.~Seibold, Mol.~Phys. (2019), DOI: 10.1080/00268976.2019.1649493.
}
{\color{black}\bibitem{GJ} N.~Gr{\o}nbech-Jensen, Mol. ~Phys. (2019), DOI: 10.1080/00268976.2019.1662506. arXiv:1909.04380.}

\bibitem{GreenKubo} M.~S.~Green,
Chem. Phys. {\bf 22}, 398 (1954); R.~Kubo,
J. Phys. Soc. Jpn. {\bf 12}, 570 (1957).
\end{thebibliography}
\end{document}